\documentclass{article}
\usepackage{graphics}
\usepackage{graphicx}
\usepackage{cmap}
\usepackage{hyperref}
\usepackage[margin=0.5in]{geometry}
\usepackage{setspace}
\usepackage{amsmath}
\usepackage{color}
\usepackage{upgreek}
\usepackage{authblk}
\usepackage{hyperref}

\title{fastbmRAG: A Fast Graph-Based RAG Framework for Efficient Processing of Large-Scale Biomedical Literature}
\date{}
\author[1]{Guofeng Meng}
\author[1]{Li Shen}
\author[1]{Qiuyan Zhong}
\author[1]{Wei Wang}
\author[1]{Haizhou Zhang}
\author[1]{Xiaozhen Wang}
\affil[1]{Changchun GeneScience Pharmaceuticals Co., Ltd. Shanghai}

\begin{document}
\maketitle

\begin{abstract}
Large language models (LLMs) are rapidly transforming various domains, including biomedicine and healthcare, and demonstrate remarkable potential from scientific research to new drug discovery. Graph-based retrieval-augmented generation (RAG) systems, as a useful application of LLMs, can improve contextual reasoning through structured entity and relationship identification from long-context knowledge, e.g. biomedical literature. Even though many advantages over naive RAGs, most of graph-based RAGs are computationally intensive, which limits their application to large-scale dataset. To address this issue, we introduce fastbmRAG, an \textbf{fast} graph-based \textbf{RAG}  optimized for \textbf{biomedical} literature. Utilizing well organized structure of biomedical papers, fastbmRAG divides the construction of knowledge graph into two stages, first drafting graphs using abstracts; and second, refining them using main texts guided by vector-based entity linking, which minimizes redundancy and computational load. Our evaluations demonstrate that fastbmRAG is over 10x faster than existing graph-RAG tools and achieve superior coverage and accuracy to input knowledge. FastbmRAG provides a fast solution for quickly understanding, summarizing, and answering questions about biomedical literature on a large scale. FastbmRAG is public available in \url{https://github.com/menggf/fastbmRAG}.

\end{abstract}

\section{Introduction}

Large language models (LLMs), such as GPT-4, have demonstrated their remarkable capabilities in understanding and generating natural language with human-like performance \cite{OpenAI2023}. These models are extensively trained on massive amounts of knowledge and are capable to generate expert-like responses in various domains \cite{Ilani2023}. Now, LLMs has been widely used in biomedical research by enhancing data analysis, automating literature reviews, and improving clinical decision-making \cite{Ahmed2023,Lu2024}.  This significantly accelerates the efficiency of researchers to acquire interest knowledge \cite{Yu2024}. However, LLMs are also challenged by incomplete or even invalid responses, known as the "hallucination" due to the biases inherent in their training data \cite{Farquhar2024, Xu2024, Yao2023}.

Retrieval-augmented generation (RAG) is a technique that enables LLMs to retrieve and incorporate new information into their responses \cite{Lewis2020}. RAG systems retrieve relevant documents or messages from some source data instead of requiring the model to memorize all knowledge, allowing them to generate more grounded and accurate responses \cite{Zhao2024}. This is especially useful to improve the performance of LLMs to handle domain-specific knowledge \cite{Lakatos2025,Fan2024}. However, performance of naive RAG approaches typically relies on proper chunk partitioning of documents and effective retrieval of related chunks according user's queries \cite{Cheng2025,Lima2024}. When it comes to large-scale domain-specific knowledge, naive RAG often retrieves irrelevant or low-quality data chunks (low precision) or fails to retrieve all relevant information (low recall) \cite{Laban2024,Li2025}.  This limits the application of naive RAG systems to integrate too much knowledge. In biomedical domain, there are millions of published literatures, which usually contain large amounts of fragmented information. Comprehensive integration is more necessary than simple retrieval. 

Recent advances in graph-based retrieval augmented with generation (RAG), such as GraphRAG \cite{Edge2024} and LightRAG \cite{Guo2024}, have introduced graph-based representations to organize the entities (e.g., genes or diseases) and their relationships. These systems use knowledge graphs to improve contextual understanding and reasoning. This results in more comprehensive, accurate, and explainable AI responses, especially for complex queries and the domains with intricate relationships \cite{Dong2024, Zhu2025}. Compared to naive RAG, graph-based RAG can integrate information from multiple documents seamlessly and uncover hidden insights and connections among them \cite{Delile2024}. LightRAG, a lightweight model, has been optimized for retrieval accuracy and efficiency and achieves a better performance. However, LightRAG still faces limitations when scaling to large biomedical data. The graph construction and inference steps are often computationally expensive and not optimized for domain-specific document structures \cite{Zhang2025}. This limits the application of graph-based RAG tools in scientific research.

To address these challenges, we present a \textbf{fast} graph-based \textbf{RAG} tool designed for processing large-scale \textbf{biomedical} literature, named as \textbf{fastbmRAG}. It utilizes  a similar framework of LightRAG by incorporates graph structures into text indexing and retrieval processes. The novel concept of fastbmRAG is to construct knowledge graphs in two steps: first, building a draft knowledge graph using only abstracts, and second, refining the relationship description of graph using both the entity relationships of the draft graph and the main texts of scientific papers. In the latter step, a vector-based search links entities in the main text, reducing the computational requirements and keeping the input chunk size within a proper range. FastbmRAG greatly reduces redundancy and accelerates information extraction in biomedical paper processing. Our evaluation shows that fastbmRAG is more than 10 times faster than LightRAG. using large-scale disease-related papers, fastbmRAG generate outputs with better coverage to disease knowledge and better description accuracy of related biomedical papers. Overall, fastbmRAG offers a fast solution for quickly understanding, summarizing, and answering questions about biomedical literature on a large scale.

\section{Methods}

\subsection{Disease-related papers}
We searched the PubMed database for disease names using queries such as "endometriosis [title/abstract] AND free full text [sb]" and downloaded the papers' PMIDs. We downloaded the full texts of these papers from the Europe PMC FTP site in XML format. We then processed the XML files using the tidyPMC package. We extracted the sections of 'title', 'authors', 'journal', 'paper ID', 'year', 'abstract', and 'main text'. Here, 'paper ID' can be a PMC ID or a PMID, and 'main text' may contain different sections. We stored this information in a Pandas DataFrame, with each paper as a row. The data frame containing all the papers is written to a text file.

\subsection{LLM model}
Under the default settings, fastbmRAG uses Ollama (v0.7.1) as the LLM backend. The Phi4 (16B) model, developed by Microsoft Research, was used to extract entities and their relationships from the input text using prompts. The 'mxbai-embed-large' embedding model was used to generate the embedded vector of the main text chunks or queries. Vector similarity was calculated using cosine similarity.

\subsection{Implement of fastbmrag and LightRAG}

Two models were implemented on an Ubuntu server with one RTX 4090 GPU (24 GB) and 1 TB of ROM. LightRAG (v1.3.8) was cloned from GitHub and installed locally. The demo script in the example directory was modified for ollama to ensure that LightRAG uses the same LLM and embedding model as fastbmRAG. The other parameters are set as the default. The main texts of all the papers were read into a Python DataFrame, and then they were indexed using the 'insert' function.

\subsection{Knowledge graph and vector database}

The knowledge graph is generated through the retrieval of entities and relationships from scientific papers. All edges are directional, with a source node and a target node. The source and target nodes are the source and target entities generated by the LLM. Each node has an 'entity\_type' attribute, such as 'gene,' 'disease,' 'mutation,' 'drug,' and other types. Node weights are calculated based on the nodes' degree in the knowledge graph and normalized according to the total number of nodes. Edges have two attributes: 'relationship\_type' and 'relationship\_description'. The 'relationship\_type' attribute is used during global queries, while the 'relationship\_description' attribute records the most important information about entity pairs. The contents of the "relationship\_description" attribute are embedded into vectors and used for similarity searching of queries. Edge weights are calculated based on node weights. By default, edge weight is calculated as the minimum weight of the source and target nodes. The edge weight indicates the confidence in the entity relationship. There are also other edge attributes, e.g., paper\_id, journal, authors, and other related information. The knowledge graph is stored in a vector database, e.g., Qdrant. 

\subsection{Query the knowledge graph}

Unlike other graph-based RAGs, fastbmRAG does not rely on users' specifications for global or local retrieval. Instead, our tool uses an LLM to parse the input query and extract information such as gene or disease names and potential output types. These correspond to the source or target node names and the 'relationship\_type' edge attribute in the knowledge graph. These can then be used to filter out irrelevant edges. This helps control the scale of the subgraph for semantic searches. Additionally, due to the existence of numerous unvalidated relationships, users can optionally set a threshold for edge weights to further filter low-confidence edges. This can further reduce the scale of the subgraph.

The query is performed by searching for vector similarity between the embedded query vector and the database records for the 'relationship\_description' of the edges. To ensure sufficient coverage of scientific papers, fastbmRAG allows the selection of an unlimited number of matched descriptions or papers. The outputs from each matched description are summarized into one integrated answer.

\subsection{Answer generation and quality assessment}

We applied fastbmRAG to real scientific papers and evaluated its output by manually checking the quality of the retrieval and generation. As described in the previous section, we used 400 open-access papers related to diseases as test dataset. We built an index with the default settings and queried it with five questions that represented some common queries in scientific research. For endometriosis, these questions are: (1) Which genes are up- or down-regulated in endometriosis? (2) Which pathways contribute to endometriosis? (3) Which mutations contribute to endometriosis? (4) What is the role of IL1B in endometriosis? (5) What is the causal mechanism of endometriosis?  The answers generated by fastbmRAG and other LLM models are evaluated by manually checking them in the original paper.

\subsection{Software and availability}
fastbmRAG is implemented in Python 3.11 or higher. There are two modes: 'update' and 'query'. The 'update' mode is used to create an new collection or add new documents to an existing collection of vector database. The documents should be a text file in CSV format. It should have at least three columns: 'abstract', 'main\_text' and 'paper\_id'. If there are more columns, they are used for additional information. Each element of the 'main\_text' column should be either a list of strings in the format '[str1, str2]' or a string separated by '$\setminus$n'. 'paper\_id' is a unique ID for each paper. If a paper ID exists in the collection, the corresponding paper will be ignored. To update the collection, use the following command:

\begin{verbatim}
python main.py --job update --document input.csv 
			   --collection_name collection_name 
			   --working_dir directory_path
\end{verbatim}

Here, ‘collection\_name’ and ‘working\_dir’ specify the collection name and directory to store collection.

Another mode is ‘query’. It is used to query the collection.

\begin{verbatim}
python main.py --job query --collection_name collection_name 
	           --working_dir directory_path 
	           --question 'your question'
\end{verbatim}

The source code of fastbmRAG is public available in \url{https://github.com/menggf/fastbmRAG}

\section{Results}
\subsection{Workflow of fastbmRAG}

\begin{figure}
	\centering
	\includegraphics[width=0.8\textwidth]{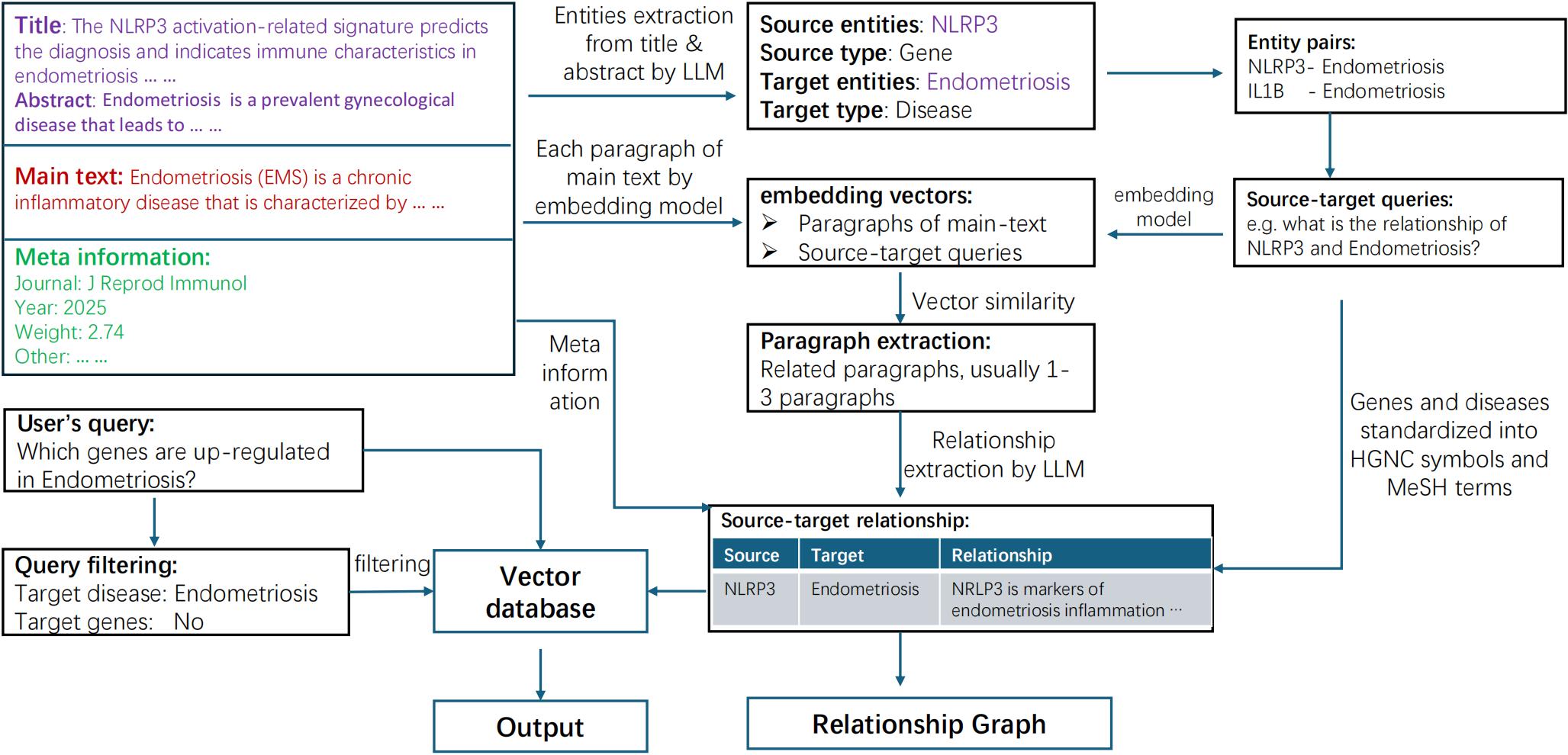}
	\caption{The flowchart of fastbmRAG to build knowledge graph}
	\label{fig1}
\end{figure}

Unlike internet documents, scientific papers are usually well-structured and include sections such as abstract, introduction, methods, results, and discussion. The abstract provides a brief yet comprehensive overview of the entire paper, while the main text has detailed yet redundant information about the key findings. FastbmRAG leverages these features of scientific papers to construct an entity-relationship knowledge graph in two steps. Figure \ref{fig1} describes the flowchart of the entire pipeline. In step one, fastbmRAG uses a similar strategy of LightRAG by incorporates graph structures into text indexing processes by entity and relationship extraction. In detail, fastbmRAG uses an LLM to process the abstracts of scientific papers and extract entities, which can be genes, diseases, drugs, animal models, mutations, pathways, or other related types. Based on their role, these entities can be either source or target entities, constituting the source or target nodes of the knowledge graph. LLM summarizes the descriptions of the relationships between source and target entities according to the knowledge in the abstracts. However, these descriptions may be concise and lack details. To refine these descriptions, fastbmRAG first generates a general query for each pair of source and target entities. In Step 2, the main texts of the scientific papers are splitted into paragraphs or fixed-size chunks. If a fixed size is set, overlapping chunks are necessary. FastbmRAG searches the query question against the embedded main text to find chunks related to the source and target entities. The LLM refines the descriptions of entity pairs using the selected chunks.

After two steps, a knowledge graph is constructed, including entities and their relationships. This graph has the following information: source entity, source entity type, target entity, target entity type, relationship description, relationship type, and additional information such as journal, authors, paper ID, publication year, and weights. To facilitate graph query, entity names should be standardized. Gene names are transformed into Human Gene Nomenclature Committee (HGNC) symbols, and disease names are transformed into Medical Subject Headings (MeSH) terms. This step is usually performed by an LLM. The knowledge graph is stored in a vector database for querying.

During the query stage, the LLM processes the question to extract the query gene, query disease, and output types. This information is then used as filters to improve the precision of the output and reduce the scale of the information graph. The query vector is then searched for relationship descriptions of the entities to retrieve related answers from related papers. Finally, the LLM summarizes these answers into a final output.

\subsection{Efficiency evaluation for indexing biomedical literature}

The primary objective of fastbmRAG is to enhance the efficiency of graph-based RAG systems for indexing large-scale biomedical literature. To evaluate its effectiveness, we collected 400 open-access papers on various diseases from the PubMed database. We built a knowledge graph using the following Python commands:
\begin{verbatim}
>import fastbmrag.fastbmrag as fastbmrag
>rag=fastbmrag.RAG(working_dir="endometriosis", collection_name="endometriosis")
>rag.insert_paper(documents)
\end{verbatim}

Here, the 'documents' is a pandas dataframe with column of 'abstract', 'main\_text', 'paper\_id', 'article\_title', 'year', 'author', 'journal', 'weight'.

We used fastbmRAG to index 400 exemplary full-text research papers on five diseases: endometriosis, psoriatic arthritis, Parkinson's disease, ovarian cancer, and COVID-19. Table \ref{table1} shows the evaluation results. On average, fastbmRAG indexed 400 papers in 1.78 hours, or about 16 seconds per paper, on a Linux server with one RTX 4090 GPU. For comparison, LightRag, a popular, general-purpose graph-based RAG model, was implemented to process the same papers using the same LLM and embedding models on the same server. With the default settings, LightRag took approximately 20.54 hours to process 400 papers, making fastbmRAG over 10 times faster. These results suggest that fastbmRAG is more efficient at indexing biomedical literature.

\begin{table}
\centering
\caption{Time consumption of fastbmRAG and LightRAG in indexing biomedical papers \label{table1}}
\begin{tabular}{cccc}
\hline
Disease$\ast$ &	No. selected Papers &	fastbmRAG &	LightRAG \\ \hline
endometriosis &	400	& 1.8 h	& 18.6 h \\
psoriatic arthritis &	400 &	1.6 h &	18.1 h \\
Parkinson's disease &	400	& 2.1 h	& 25.3 h \\
Ovarian cancer &	400 & 1.9 h & 22.2 h \\
COVID-19	& 400 & 1.5 h & 18.5 h \\ \hline
Average & 400 &	1.78 h & 20.54 h \\ \hline

\hline
\end{tabular}

{ \tiny $\ast$Evaluation is repeated for three times and the average computational time is reported.}

\end{table}

\subsection{Outcome evaluation}
\begin{figure}
	\centering
	\includegraphics[width=0.9\textwidth]{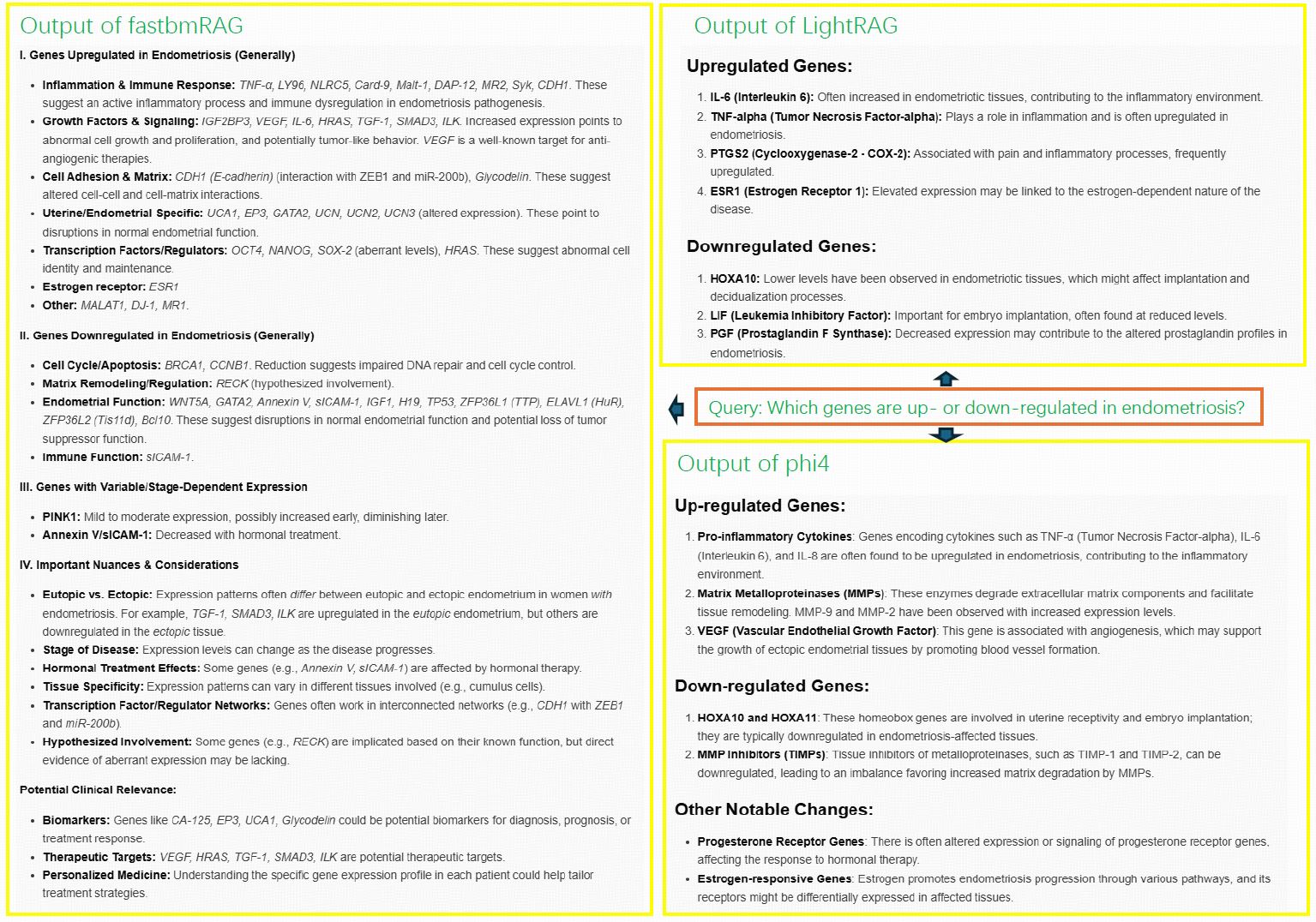}
	\caption{Outputs of fastbmRAG, LightRAG and Phi4 model}
	\label{fig2}
\end{figure}

We evaluated FastbmRAG using 7,517 open-access, full-text papers related to endometriosis. After indexing, we used it to answer some endometriosis-related questions and evaluated the coverage and accuracy of the results. The first question was, "Which genes are up- or down-regulated in endometriosis?" Figures \ref{fig2} and Table S1 show the output of fastbmRAG under different queries. Compared to the outputs of LightRag and the phi4 model, fastbmRAG collected more up- or down-regulated genes and classified them into different categories. FastbmRAG also recognized the importance of other covariates (e.g., tissue regions, hormonal treatment, menstrual stage, and other factors that affect gene expression) and discussed them in its output.  We notices some difference in output of fastbmRAG and LightRAG. For examples, LightRag reported down-regulation of PTGS2. Our manual investigation found that only 9 open-access full-text papers reported the association of PTGS2 with endometriosis and no paper reported the differential expression status of PTGS2 under disease status. This is the reason why fastbmRAG fails to identify some genes. We manually verified the output genes of fastbmRAG and found that the reported up- or down-regulated genes were all generated according to the source papers. We performed the same evaluation with other questions, and fastbmRAG always generated more detailed outputs. Overall, our evaluation suggests that fastbmRAG performs well in generating answers according to the user's query with good coverage and accuracy.

\section{Conclusion and Discussion}

FastbmRAG is a new graph-based RAG tool, characterized by (a) fast-speed to process source dataset and (b) specifically designed for biomedical papers. Unlike GraphRAG and LightRAG, fastbmRAG is specifically designed for biomedical literature. It utilizes the well-organized structure of scientific papers, such as the abstract and main text. FastbmRAG use a similar framework of LightRAG by incorporates graph structures into text indexing and retrieval processes. It builds a knowledge graph in two steps. First, it extracts entities from the abstract text. Then, it builds a knowledge graph using only the abstract text, which is usually shorter than the main text but contains all the information from the main text. The knowledge graph generated using abstracts usually contains the key entities and entity relationships but has fewer detailed descriptions of the entity relationships. In the next step, fastbmRAG refines the entity relationships using the main text. Using semantic searching, fastbmRAG extracts related paragraphs from the main text — usually two to three paragraphs — and refines the relationship descriptions of the existing knowledge graph with these paragraphs. This makes fastbmRAG more efficient at building knowledge graphs than general graph-based RAGs. When applied to real data, our evaluation suggested that fastbmRAG indexed scientific papers about ten times faster than LightRAG.

Biomedical papers  always have limited number of entities and types. In biomedical papers, genes and diseases are key entities, and their names are standardized as specific terms (e.g., HGNC gene symbols and Disease Ontology terms). Transforming the alias names into standard names allows fastbmRAG to filter the knowledge graph based on the exact matching of disease and gene names. This greatly increases querying efficiency and generates more precise responses to users' questions. Our evaluation also shows that fastbmRAG can capture more precise information from the input papers according to the user's query.

FastbmRAG also has some limitations. Although the settings make fastbmRAG more efficient to process biomedical papers, it may miss the knowledge that only exists in the in main texts or that authors did not give enough description. One example is PTGS2 that has been discussed in above section. Current version of fastbmRAG still needs many improvement, e.g. integrating reasoning in the outputs to generate more confident results.

\section{Acknowledgements}
This work is supported by Changchun GeneScience Pharmaceuticals Co., Ltd..

\end{document}